\documentclass[conference]{IEEEtran}
\IEEEoverridecommandlockouts 

\usepackage{cite} 
\usepackage{amsmath,amssymb,amsfonts} 
\usepackage{algorithmic} 
\usepackage{graphicx} 
\usepackage{textcomp} 
\usepackage{xcolor} 
\usepackage{url} 
\usepackage{booktabs} 
\usepackage{makecell} 
\usepackage{enumitem} 
\usepackage{microtype} 

\usepackage{hyperref}
\usepackage{graphicx}
\usepackage{float} 

\newcommand{\orcidicon}[1]{%
    \href{https://orcid.org/#1}{%
        \includegraphics[width=10pt]{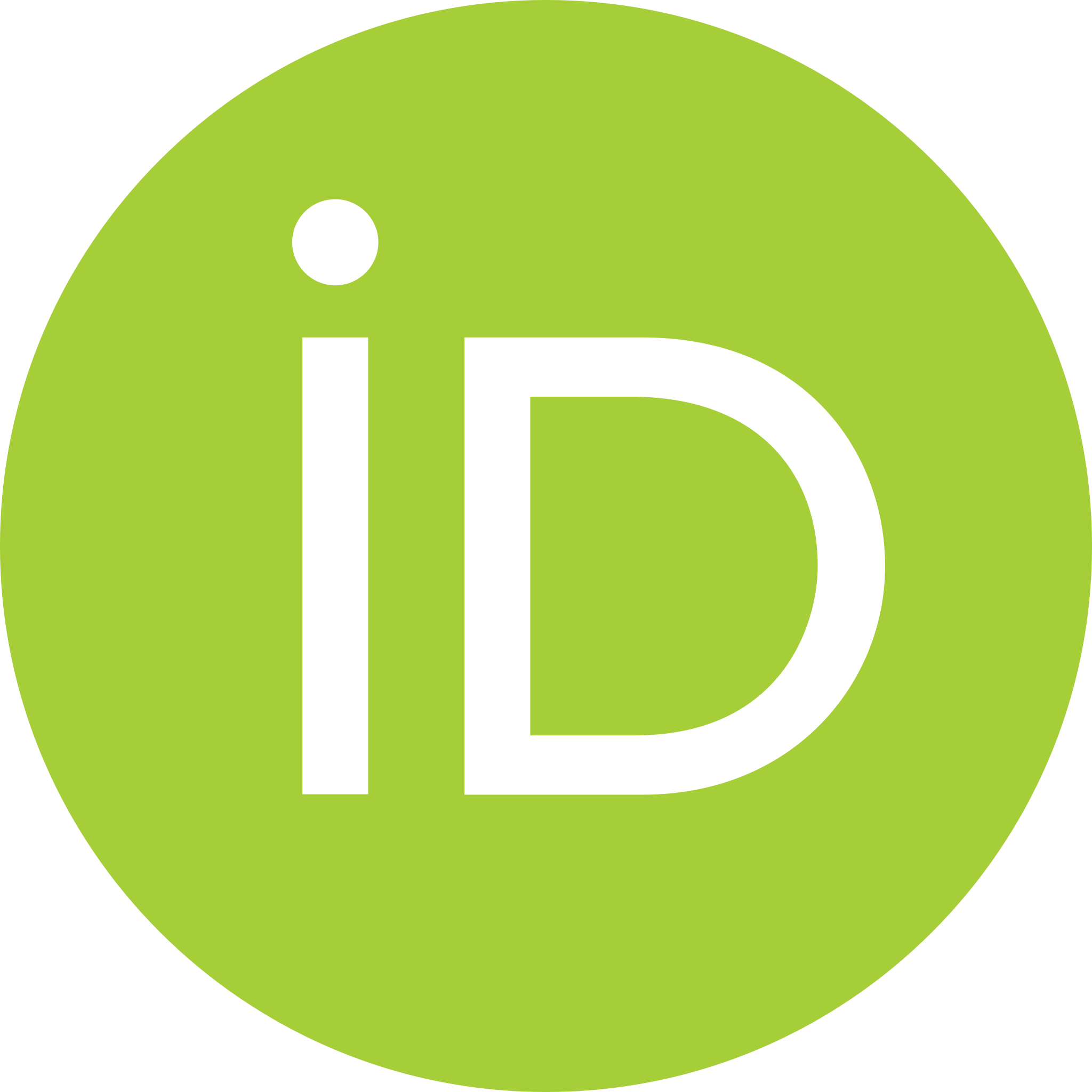}%
    }%
}

\def\BibTeX{{\rm B\kern-.05em{\sc i\kern-.025em b}\kern-.08em
    T\kern-.1667em\lower.7ex\hbox{E}\kern-.125emX}}

\setlist[itemize]{leftmargin=*, topsep=0pt, partopsep=0pt, parsep=0pt}

\begin{document}

\title{Building A Secure Agentic AI Application Leveraging Google's A2A Protocol}


\author{
\IEEEauthorblockN{Idan Habler\textsuperscript{1}\thanks{\textsuperscript{1}These authors contributed equally to this work.} \textsuperscript{2}\thanks{\textsuperscript{2}This work is not related to the author’s position at Intuit}}
\IEEEauthorblockA{\textit{Adversarial AI Security reSearch (A2RS)} \\
\textit{Intuit} \\
idan\_habler@intuit.com \textsuperscript{\orcidicon{0000-0003-3423-5927}
}
}
\and
\IEEEauthorblockN{Ken Huang\textsuperscript{1} \textsuperscript{3}\thanks{\textsuperscript{3}This work is not related to the author’s position at DistributedApp.ai}} 
\IEEEauthorblockA{\textit{Agentic AI Security} \\
\textit{DistributedApps.ai} \\
ken.huang@distributedapps.ai \textsuperscript{\orcidicon{0009-0004-6502-3673}
}
}
\\ 
\IEEEauthorblockN{Prashant Kulkarni\textsuperscript{1} \textsuperscript{5}\thanks{\textsuperscript{5}This work is not related to the author’s position at Google}}
\IEEEauthorblockA{\textit{Google Cloud} \\
\textit{Google} \\
pskulkarni@google.com \textsuperscript{\orcidicon{0009-0004-2344-4840}
}
}
\and
\IEEEauthorblockN{Vineeth Sai Narajala\textsuperscript{1} \textsuperscript{4}\thanks{\textsuperscript{4}This work is not related to the author’s position at Amazon Web Services.}}
\IEEEauthorblockA{\textit{Proactive Security} \\
\textit{Amazon Web Services} \\
vineesa@amazon.com \textsuperscript{\orcidicon{0009-0007-4553-9930}
}
}
} 

\maketitle 
\begin{abstract}
As Agentic AI systems evolve from basic workflows to complex multi-agent collaboration, robust protocols such as Google's Agent2Agent (A2A) become essential enablers. To foster secure adoption and ensure the reliability of these complex interactions, understanding the secure implementation of A2A is essential. This paper addresses this goal by providing a comprehensive security analysis centered on the A2A protocol. We examine its fundamental elements and operational dynamics, situating it within the framework of agent communication development. Utilizing the MAESTRO framework, specifically designed for AI risks, we apply proactive threat modeling to assess potential security issues in A2A deployments, focusing on aspects such as Agent Card management, task execution integrity, and authentication methodologies. 

Based on these insights, we recommend practical secure development methodologies and architectural best practices designed to build resilient and effective A2A systems. Our analysis also explores how the synergy between A2A and the Model Context Protocol (MCP) can further enhance secure interoperability. This paper equips developers and architects with the knowledge and practical guidance needed to confidently leverage the A2A protocol for building robust and secure next-generation agentic applications.

\end{abstract}



\section{Introduction}
The emergence of intelligent, autonomous agents marks a pivotal shift in how AI systems are developed, deployed, and scaled. As these agents increasingly interact across organizational and technological boundaries, the need for secure, interoperable communication becomes critical. This paper begins by exploring the foundation of this transformation: the rise of Agentic AI and the protocols that enable it.
\subsection{The Rise of Agentic AI and the Need for Secure Interoperability}
As artificial intelligence systems evolve from isolated, task-specific models to dynamic, multi-agent ecosystems, we are witnessing the emergence of Agentic AI—intelligent agents capable of autonomous decision-making, tool use, and collaboration with other agents and humans. These agents do not merely respond to prompts; they initiate actions, delegate subtasks, coordinate with peers, and adapt to new goals in real time. From AI research assistants that plan literature reviews to supply chain agents negotiating logistics across organizations, agentic AI is rapidly becoming the backbone of next-generation intelligent applications.
However, as these agents interact and compose workflows across organizational, geographical, and trust boundaries, the need for secure, standardized interoperability becomes paramount. Without a shared protocol for identity, authentication, task exchange, and auditability, agent interactions are prone to fragmentation, redundancy, and most critically—security vulnerabilities. Threats such as impersonation, data exfiltration \cite{llm_genai_security_2025}, task tampering, and unauthorized privilege escalation can quickly arise in loosely governed agent ecosystems.
To address this, secure interoperability protocols like Google’s Agent-to-Agent (A2A) specification offer a promising foundation. A2A provides a declarative, identity-aware framework for enabling structured, secure communication between agents—whether human-authored or AI-powered. 
Such protocols enable agents to share descriptive information about their capabilities, which is essential for facilitating effective interaction, discovery, and interoperability. This exchanged information, especially when received from potentially untrusted peers, must be handled with care and rigorously validated to prevent manipulation techniques like prompt injections. 
Realizing the full potential of agentic AI will depend not only on such protocol standards, but on robust implementations, rigorous threat modeling, and continuous security adaptation. To aid developers in building secure systems, we also offer a repository containing secure A2A coding examples \footnote{\url{https://github.com/kenhuangus/a2a-secure-coding-examples} \label{fn:gh_repo}}.

This paper explores the security architecture of A2A, identifies critical risks through the MAESTRO threat modeling lens ~\cite{Huang2025Maestro,huang2025threat} and proposes mitigation strategies and implementation best practices to ensure that agentic systems remain not just intelligent—but trustworthy by design.

\subsection{Google A2A: A Foundational Protocol and its Context}
The Agent-to-Agent (A2A) protocol, introduced by Google, represents a significant step forward in enabling structured, secure, and interoperable communication between autonomous agents. Designed with composability and trust in mind, A2A allows agents to discover each other via standardized AgentCards, authenticate using modern cryptographic protocols, and exchange tasks in a declarative, auditable manner. Its architecture reflects the growing demand for modular AI systems that can scale across organizations, tools, and domains—while remaining adaptable to both human and machine-driven workflows. A2A’s emergence is timely, as the AI ecosystem increasingly shifts toward open-ended, agent-driven applications in areas like research, enterprise automation, cybersecurity, and scientific collaboration. Positioned at the intersection of protocol engineering and AI orchestration, A2A provides the plumbing needed to build reliable, multi-agent ecosystems—where interoperability and security are not afterthoughts, but built-in foundations.

\subsection{Paper Objectives: Analyzing A2A, Proposing Enhancements, and Guiding Secure Implementation}
This paper sets out to examine the Agent-to-Agent (A2A) protocol in both its theoretical design and practical deployment, with the overarching goal of enabling secure and trustworthy agentic AI systems. First, we analyze the A2A protocol through the lens of the MAESTRO threat modeling framework, identifying a comprehensive set of risks that emerge in multi-agent environments—including spoofing, task replay, privilege escalation, and prompt injection. Building on this analysis, we propose targeted security enhancements to strengthen the protocol and its implementations, ranging from cryptographic controls and schema enforcement to secure session handling and fine-grained authorization \cite{mas_threat_model_2025}. Finally, the paper outlines a set of implementation best practices for building hardened A2A servers and clients, offering guidance on secure communication, authentication workflows, logging, and abuse prevention. By bridging the gap between protocol specification and real-world deployment, this paper aims to equip developers, researchers, and architects with actionable insights for designing secure-by-default agentic ecosystems.

\section{Literature Review}

\subsection{Agent to Agent Communication Literature Review}
Agent-to-agent communication is fundamental to multi-agent systems (MAS), enabling autonomous entities to coordinate and collaborate to solve complex problems \cite{finin1994kqml}. Efficient and secure communication protocols are critical for sophisticated agent interactions across heterogeneous platforms. This review synthesizes the historical development, protocols, theoretical foundations, applications, challenges, and future directions of agent communication, with a focus on secure A2A multi-agent systems.

Early agent communication languages (ACLs) emerged from the Knowledge Sharing Effort (KSE) in the early 1990s \cite{labrou2000history}. KQML, the pioneering ACL, defined a multi-layered architecture including content, message, and communication layers using speech act theory through performatives to specify message intentions \cite{finin1994kqml}. Building upon KQML, the FIPA ACL sought more rigor with a semantic framework based on modal logic, feasibility preconditions, and rational effects \cite{fipa_acl, chaibdraa2002trends}. These ACLs are based on speech act theory, formal semantics, and the Belief-Desire-Intention (BDI) model \cite{gaudou2006new}.

Key protocols include KQML with its LISP-like syntax, and FIPA ACL which refines message structure and semantics with defined preconditions and outcomes \cite{finin1994kqml}. Protocols define interaction sequences between agents, including request, query, contract net, auction, and subscription \cite{fipa_acl}. Recent efforts like Google's A2A protocol aim to enable seamless interoperability between AI agents across different frameworks by standardizing task negotiation, capability sharing, and secure communication \cite{googleblog2025a2a}.

Communication architectures in MAS can be direct (full, partial, or dynamic networks) or mediated (facilitator-based, blackboard systems, or global control) \cite{dorri2018multi}. Multi-agent reinforcement learning (MARL) allows agents to learn when, how, and what to communicate using differentiable, reinforcement, supervised, or regularized communication learning methods \cite{zhu2024survey}.

Applications of agent communication range across enterprise knowledge management, customer service, supply chain management, healthcare, smart grids, autonomous vehicles, e-commerce, and disaster response \cite{dorri2018multi, fipa_acl}. Security, trust, and the potential manipulation of feedback loops present inherent challenges. Integration with existing standards, scaling communication for efficiency, and handling non-stationarity in learning processes are open research areas \cite{labrou2000history, googleblog2024placeholder, fipa_acl}. 

Future research should focus on multimodal communication, structured communication networks, robust centralized components, interpretable emergent languages, and integration with LLMs. Initiatives like A2A and MCP are aiming for protocol convergence for standardized agent communication, although many questions regarding how this may emerge are still unanswered. Effective communication remains crucial for the success of agent systems. Security must be considered throughout their evolution.

\subsection{Traditional PKI and Transport Security}
Early implementations of secure agentic systems leveraged public key infrastructure (PKI), with each agent provisioned a certificate and private key signed by a trusted certificate authority. This approach, frequently combined with SSL/TLS at the transport layer, guaranteed encrypted channels for agent communication \cite{bellare1995optimal}. While PKI solutions remain fundamental to securing multi-agent messages, they often required complex certificate management, especially as agent populations grew or changed dynamically. Agent platforms such as JADE (Java Agent DEvelopment Framework) attempted to simplify this via security extensions that integrated PKI-based authentication \cite{jade_security}.

\subsection{SOAP, WS-Security, and Enterprise Service Buses}
Prior to the widespread adoption of RESTful services, enterprise-grade agentic applications often employed SOAP as the foundational protocol. The WS-Security suite \cite{oasis2004wssecurity} provided message-level encryption, signatures, and token-based authentication for agent-to-service or agent-to-agent interactions. These frameworks were robust but verbose; they were well-suited to closed enterprise environments yet burdensome for lightweight agentic systems or real-time AI workflows. Still, they established fundamental principles of end-to-end message confidentiality, integrity, and non-repudiation at a time when microservices and ephemeral containers were not standard practice.

\subsection{OAuth 2.0, JSON Web Tokens (JWT), and Service Accounts}
With the shift to cloud and user-centric web applications, OAuth 2.0 became the de facto framework for delegated authorization \cite{hardt2012oauth}. However, OAuth 2.0’s flows primarily focus on interactive user login. This made it challenging for autonomous agents to securely obtain and renew tokens without user intervention. JSON Web Tokens (JWT) \cite{jones2015jwt} introduced a lightweight format for securely transmitting agent identity and privileges. Many practitioners adopted service accounts or “robot accounts” with static credentials to simulate user roles in an agent scenario, though this introduced key-rotation and secret-management complexities.

\subsection{Zero-Trust Architectures and BeyondCorp}
As multi-agent systems expanded across hybrid cloud environments, Zero-Trust principles emphasized continuous validation of identity and authorization, even for internal requests \cite{rose2020zero}. Solutions like Google’s BeyondCorp architecture applied these principles, requiring every agent or service to authenticate with strong, verifiable credentials irrespective of network location \cite{ward2019adopting}. This paved the way for “never trust, always verify” policies, which align closely with the needs of autonomous AI agents that may operate across heterogeneous network boundaries.

\subsection{Limitations of Historical Approaches}
The historical approaches to securing agentic AI systems shared several common limitations:
\begin{itemize}
    \item \textbf{Manual Credential Management:} Significant manual effort was required to provision, rotate, and revoke agent credentials.
    \item \textbf{Adapting User-Centric Models:} Substantial work was needed to adapt inherently user-centric security workflows to machine-only environments.
    \item \textbf{Fragmented Implementations:} The lack of uniform standards or cloud-native support for fully autonomous agents resulted in inconsistent security implementations.
    \item \textbf{Scaling Challenges:} As agent populations grew, the operational complexity of maintaining secure communication increased exponentially.
\end{itemize}
These early innovations and their limitations highlighted the fundamental importance of establishing robust identity and trust mechanisms for distributed AI, while also underscoring the need for simpler, more automated credential exchange processes specifically designed for autonomous agent interactions.

\section{Understanding Google's A2A Protocol}

\subsection{Protocol Overview}
The A2A protocol facilitates communication between two primary types of agents:
\begin{itemize}
    \item \textbf{Client Agent:} Responsible for formulating and communicating tasks.
    \item \textbf{Remote Agent:} Responsible for acting on those tasks to provide information or take action.
\end{itemize}
Built on established web standards including HTTP, JSON-RPC, and Server-Sent Events (SSE), A2A prioritizes compatibility with existing systems while maintaining a security-first approach. Its design follows several key principles:
\begin{itemize}
    \item \textbf{Agentic-first:} Agents operate independently and communicate explicitly to exchange information, without shared memory or tools by default.
    \item \textbf{Standards-compliant:} Utilization of widely adopted web technologies minimizes developer friction.
    \item \textbf{Secure by default:} Integrated authentication and authorization measures safeguard sensitive data and transactions.
\end{itemize}

\subsection{Communication Flow}
The A2A protocol defines a structured communication flow between agents that consists of several stages:
\begin{enumerate}
    \item \textbf{Discovery:} The client fetches the target agent's Agent Card (from \texttt{/.well-known/agent.json}) via HTTPS to learn about its capabilities, endpoint, and authentication requirements.
    \item \textbf{Initiation:} After authenticating using a method specified in the Agent Card, the client sends a JSON-RPC request over HTTPS to the server's \texttt{a2aEndpointUrl} using one of two methods:
    \begin{itemize}
        \item \texttt{tasks.send}: Used for potentially synchronous tasks or initiating tasks where immediate streaming isn't required.
        \item \texttt{tasks.sendSubscribe}: Used for long-running tasks requiring streaming updates, establishing a persistent HTTPS connection for Server-Sent Events (SSE).
    \end{itemize}
    \item \textbf{Processing \& Interaction:}
    \begin{itemize}
        \item \textit{Non-Streaming (\texttt{tasks.send}):} The server processes the task and returns the final Task object in the HTTP response.
        \item \textit{Streaming (\texttt{tasks.sendSubscribe}):} The server sends SSE messages over the persistent connection, including \texttt{TaskStatusUpdateEvent} (containing the updated Task object) and \texttt{TaskArtifactUpdateEvent} (containing generated Artifact objects).
    \end{itemize}
    \item \textbf{Input Required:} If additional input is needed, the server signals this via the response or an SSE event, and the client sends subsequent input using the same \texttt{taskId}.
    \item \textbf{Completion:} The task transitions to a terminal state (completed, failed, or canceled), communicated via the final response or an SSE event.
    \item \textbf{Push Notifications (Optional):} Servers supporting the \texttt{pushNotifications} capability can send asynchronous updates to a client-provided webhook URL, registered using \texttt{tasks.pushNotification.set}.
\end{enumerate}

\subsection{Discoverability Mechanism}
A central innovation of the A2A protocol is its robust discoverability mechanism, which enables agents to efficiently locate and leverage each other's capabilities.

\subsubsection{AgentCard Structure}
The AgentCard serves as the foundation of the A2A protocol's discoverability mechanism. This structured JSON metadata file, located at the standardized path \texttt{/.well-known/agent.json}, functions as a machine-readable "business card" describing an agent's capabilities and interfaces. The AgentCard contains:
\begin{itemize}
    \item Basic agent identification (name, version, provider).
    \item HTTPS endpoint URL for A2A communication (\texttt{a2aEndpointUrl}).
    \item Detailed function catalogs with parameter schemas.
    \item Required authentication methods using OpenAPI 3.x Security Scheme objects (e.g., \texttt{apiKey}, \texttt{http bearer}, \texttt{oauth2}, \texttt{openIdConnect}).
    \item Capability descriptions that define available functions.
    \item Hosted/DNS information detailing accessibility.
\end{itemize}
This standardized location creates a predictable discovery pattern across the entire A2A ecosystem, similar to how \texttt{robots.txt} and \texttt{sitemap.xml} function for web crawlers. The AgentCard's JSON structure ensures both human readability and machine parsability, facilitating seamless integration between disparate agent systems.

\subsection{Core Concepts}
The A2A protocol is built around several fundamental concepts that structure agent interactions:

\subsubsection{AgentCard}
As described earlier, the Agent Card is a public JSON metadata file that acts as a machine-readable business card. The quality and accuracy of Agent Cards can vary, which directly impacts discovery reliability and security posture advertisement.

\subsubsection{A2A Server}
An agent exposing an HTTPS endpoint that implements the A2A JSON-RPC methods. It receives requests, manages task execution, and sends responses/updates.

\subsubsection{A2A Client}
An application or another agent that consumes A2A services by sending JSON-RPC requests to a server's \texttt{a2aEndpointUrl}. Clients can dynamically add and remove servers after launch, offering flexibility in how capabilities are discovered and utilized during runtime.

\subsubsection{Task}
The fundamental unit of work in the A2A protocol, identified by a unique \texttt{taskId} provided by the client. Tasks progress through a defined lifecycle represented by \texttt{TaskStatus} values:
\begin{itemize}
    \item \texttt{TASK\_STATUS\_SUBMITTED}
    \item \texttt{TASK\_STATUS\_WORKING}
    \item \texttt{TASK\_STATUS\_INPUT\_REQUIRED}
    \item \texttt{TASK\_STATUS\_COMPLETED}
    \item \texttt{TASK\_STATUS\_FAILED}
    \item \texttt{TASK\_STATUS\_CANCELED}
\end{itemize}
The protocol supports both short-lived request/response style interactions and long-running asynchronous tasks, making it suitable for complex workflows that may extend over days or weeks.

\subsubsection{Message}
Represents a turn in the communication dialogue, containing a role ("user" for client-originated, "agent" for server-originated) and one or more Parts.

\subsubsection{Part}
The basic unit of content within a Message or Artifact. Defined types include:
\begin{itemize}
    \item \texttt{TextPart:} For unstructured plain text communication.
    \item \texttt{FilePart:} For transferring files, either via inline Base64 encoded \texttt{bytesContent} or a URI reference.
    \item \texttt{DataPart:} For structured JSON data, identified by a \texttt{mimeType} (e.g., \texttt{application/json}).
\end{itemize}

\subsubsection{Artifact}
Represents outputs generated by the agent during task execution (e.g., files, reports, structured data), also containing Parts.

\section{Threat Modeling Google A2A based Agentic AI Apps with MAESTRO}
This section will focus on typical agentic AI applications built using Google A2A Applications. We will use the MAESTRO threat modeling framework \cite{Huang2025Maestro} and the work documented in \cite{huang2025threat} to build application-specific threats for Agentic AI applications built using the Google A2A protocol. The next section will dive deep into strategies to mitigate the threats.
\subsection{Recap of MAESTRO Threat Modeling Methodology}
Traditional threat modeling frameworks often fall short when applied to agentic AI systems. These systems can autonomously make decisions, interact with external tools, and learn over time – capabilities that introduce unique security risks. That's why we'll use the MAESTRO framework, a seven-layer threat modeling approach specifically designed for agentic AI. MAESTRO offers a more granular and proactive methodology uniquely suited for the complexities of agentic systems like those built using A2A.
MAESTRO (Multi-Agent Environment, Security, Threat, Risk, and Outcome) provides a structured, granular, and proactive methodology for identifying, assessing, and mitigating threats across the entire agentic AI lifecycle.

MAESTRO in a Nutshell:
\begin{itemize}
    \item Extends Existing Frameworks: Builds upon established security frameworks like STRIDE, PASTA, and LINDDUN, but adds AI-specific considerations.
    \item Layered Security: Recognizes that security must be addressed at every layer of the agentic architecture.
    \item AI-Specific Threats: Focuses on the unique threats arising from AI, such as adversarial machine learning and the risks of autonomous decision-making.
    \item Risk-Based Approach: Prioritizes threats based on their likelihood and potential impact.
    \item Continuous Monitoring: Emphasizes the need for ongoing monitoring and adaptation.
\end{itemize}
The Seven Layers of MAESTRO (See Figure 1):
\begin{enumerate}
    \item Foundation Models: The core AI models (e.g., LLMs) used by the agents.
    \item Data Operations: The data used by the agents, including storage, processing, and vector embeddings.
    \item Agent Frameworks: The software frameworks and APIs that enable agent creation and interaction (like the A2A protocol).
    \item Deployment and Infrastructure: The underlying infrastructure (servers, networks, containers) that hosts the agents and API.
    \item Evaluation and Observability: The systems used to monitor, evaluate, and debug agent behavior.
    \item Security and Compliance: The security controls and compliance measures that protect the entire system.
    \item Agent Ecosystem: The environment where multiple agents interact, including marketplaces, collaborations, and potential conflicts.
\end{enumerate}
\begin{figure}
    \centering
    \includegraphics[width=0.99\linewidth]{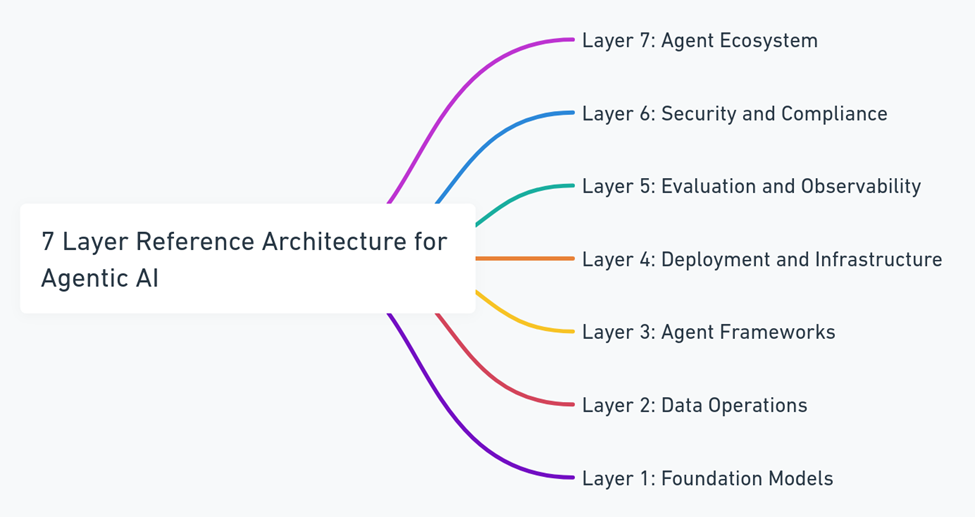}
    \caption{Maestro Architecture - 7 Layers}
    \label{fig:enter-label}
\end{figure}

\subsection{Common A2A Multi-Agent System Threats}\label{sec:threats}
\begin{figure*}
    \centering
    \includegraphics[width=1\linewidth]{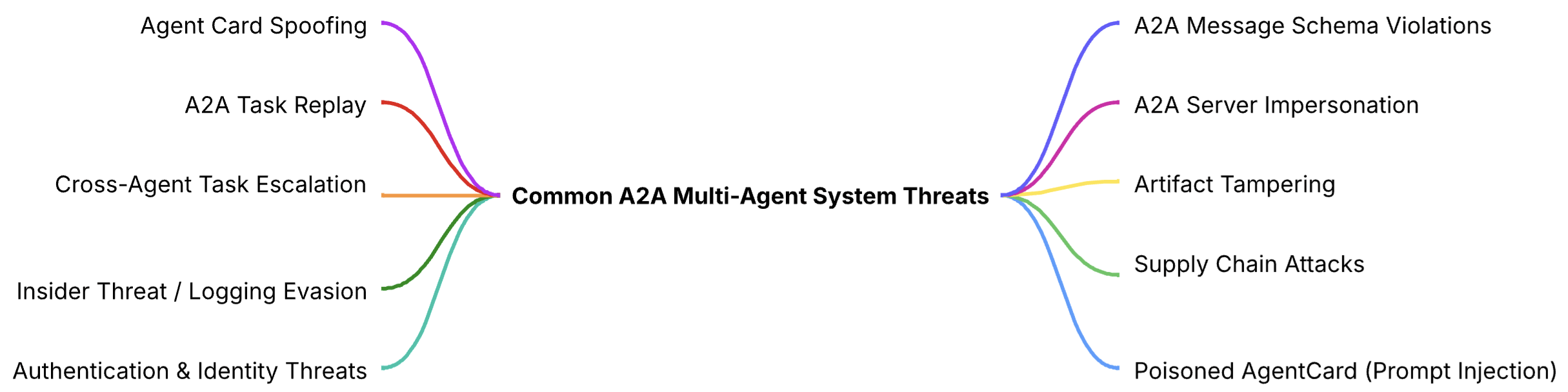}
    \caption{List of Common A2A Multi-Agent System Threats Identified by MAESTRO Threat Modeling Methodology}
    \label{fig:a2athreats}
\end{figure*}

Leveraging MAESTRO threat modeling methodology, we identified potential threats for A2A multi-agent systems, which are illustrated in Figure~\ref{fig:a2athreats} and detailed below:
\subsubsection{Agent Card Spoofing}\label{agent-card-spoofing}
\textbf{MAESTRO Layers: 3 (Agent Frameworks), 4 (Deployment \& Infrastructure)} \\
An attacker publishes a forged /.well-known/agent.json (Agent Card) at a malicious or typosquatting domain. When an A2A Client performs agent discovery, it may trust this fake card and send sensitive A2A Tasks to a rogue A2A Server. This can result in task hijacking, data exfiltration, and agent impersonation.

\subsubsection{A2A Task Replay}\label{a2a-task-replay}
\textbf{MAESTRO Layers: 3 (Agent Frameworks), 2 (Data Operations)}
If an attacker captures a valid \texttt{tasks/send} request and replays it to the A2A Server, the same A2A Task may be executed multiple times. Without replay protection, this leads to duplicate or unauthorized actions.

\subsubsection{A2A Message Schema Violation}\label{a2a-schema-violation}
\textbf{MAESTRO Layer: 2 (Data Operations)}
A malicious A2A Client may craft malformed A2A Messages or Parts to exploit weak schema validation in the A2A Server, potentially causing code injection, privilege escalation, or denial of service.

\subsubsection{A2A Server Impersonation}\label{a2a-server-impersonation}
\textbf{MAESTRO Layer: 4 (Deployment \& Infrastructure)}
Through DNS spoofing or network attacks, an adversary redirects A2A Client traffic to a fake A2A Server. The attacker can serve forged Agent Cards and Task results, undermining trust and stealing data.

\subsubsection{Cross-Agent Task Escalation}\label{cross-agent-task-esc}
\textbf{MAESTRO Layers: 7 (Agent Ecosystem), 3 (Agent Frameworks)}
A malicious agent enumerates available Agent Cards and attempts to escalate privileges by submitting A2A Tasks with forged credentials, breaching trust boundaries and accessing unauthorized data.

\subsubsection{Artifact Tampering via A2A Artifacts}\label{a2a-artifact-tampering}
\textbf{MAESTRO Layers: 2 (Data Operations), 3 (Agent Frameworks)}
Attackers intercept or modify Artifacts exchanged during A2A Task execution, injecting malicious content or corrupting results.

\subsubsection{Insider Threat/Logging Evasion via A2A Task Manipulation}\label{insider-threat}
\textbf{MAESTRO Layers: 6 (Security \& Compliance), 3 (Agent Frameworks)}
A privileged user or agent manipulates Task state transitions or disables logging on the A2A Server, concealing unauthorized actions.

\subsubsection{Supply Chain Attack via A2A Dependencies}\label{supply-chain}
\textbf{MAESTRO Layers: 4 (Deployment \& Infrastructure), 6 (Security \& Compliance)}
A compromised or vulnerable dependency in the A2A Server or Client can enable remote code execution or credential theft.

\subsubsection{Authentication \& Identity Threats}\label{auth-identity-threats}
\textbf{MAESTRO Layers: 3 (Agent Frameworks), 4 (Deployment \& Infrastructure), 6 (Security \& Compliance), 7 (Agent Ecosystem)}
A2A-based systems use OAuth/OIDC with JWT tokens for identity validation. Threats include:
\begin{itemize}
    \item Forged or stolen JWT tokens allowing unauthorized access to A2A Tasks or Agent Cards.
    \item Weak JWT validation (e.g., missing signature check, improper audience/issuer claims).
    \item Token replay or use of expired tokens.
    \item Insecure storage or transmission of tokens.
\end{itemize}

\subsubsection{Poisoned AgentCard}\label{poisioned-agetnt-card}
\textbf{MAESTRO Layers: 1 (Foundation Model), 2 (Data Operations)}
An attacker embeds malicious instructions using prompt injection techniques  within the fields of an AgentCard (e.g., AgentSkill id, name, descriptions, tags or examples). When another agent automatically ingests and processes this AgentCard data as part of its discovery or interaction flow, it may execute these hidden instructions. 
This exploits the trust placed in AgentCard content (Data Operations) and the automated processing of this content using the foundation model during its planning. As a result, the agent's goals being hijacked, sensitive data being revealed, or internal security protocols being bypassed, highlighting the need to treat AgentCard content as untrusted input requiring careful validation and sanitization.

\subsection{Additional Security Considerations}\label{threats:addition sec}
In addition to the above specific controls, we think following additional security aspects should be considered:
\begin{itemize}
    \item Supply chain dependencies must be scanned and verified. (Layers 4, 6)
    \item Incident response and recovery plans must be in place. (Layers 6, 7)
    \item Use open-source libraries for validation, authentication, and monitoring. (Layers 2, 3, 6)
    \item Prefer declarative security configurations (YAML/JSON) for infrastructure. (Layer 4)
    \item Integrate security testing (unit, integration, adversarial) into CI/CD pipelines. (Layers 3, 6)
    \item Document all agent capabilities and trust boundaries in the Agent Card. (Layers 3, 7)
\end{itemize}

\section{Case Study}
We will consider two case studies to understand the threat modeling of the agentic system.
\subsection{Case Study 1: Collaborative Document Processing}\label{case-study-1}
In this scenario, multiple A2A Clients (from different vendors) discover and interact with an enterprise A2A Server to co-edit, summarize, and review documents. Each client retrieves the Agent Card, authenticates, and launches A2A Tasks via \texttt{tasks/send}.

\begin{enumerate}
    \item \textbf{Layer 1: Foundation Models} — Prompt injection attacks can occur when adversarial input is embedded in A2A Message Parts, causing the LLM to behave unexpectedly.
    \item \textbf{Layer 2: Data Operations} — Attackers may leak sensitive data through A2A Artifacts or tamper with A2A Task state.
    \item \textbf{Layer 3: Agent Frameworks} — Vulnerable to Agent Card spoofing and replayed tasks/send requests, especially if malformed Agent Cards are accepted.
    \item \textbf{Layer 4: Deployment \& Infrastructure} — Risks include denial-of-service attacks via flooding the A2A Server with tasks, or lateral movement using compromised Agent Cards.
    \item \textbf{Layer 5: Evaluation \& Observability} — Log tampering or insufficient auditing of A2A Task transitions can let attacks go undetected.
    \item \textbf{Layer 6: Security \& Compliance} — Credential theft from Agent Cards or bypassing policies via misconfigured fields.
    \item \textbf{Layer 7: Agent Ecosystem} — Enumeration of Agent Cards and Sybil attacks with fake cards can undermine trust in federated A2A deployments.
\end{enumerate}
\subsubsection{Cross-Layer Vulnerabilities}
\begin{itemize}
    \item Compromised Agent Card enables attacker to hijack Task execution (Layer 3 $\rightarrow$ Layer 2).
    \item Weak authentication on A2A Server allows replayed Task requests (Layer 3 $\rightarrow$ Layer 6).
    \item Insufficient logging of Task state changes enables undetected attacks (Layer 5 $\rightarrow$ Layer 7).
\end{itemize}

\subsubsection{\textbf{Risk Assessment}}
\begin{itemize}
    \item Likelihood: \textbf{High} (open Agent Card discovery, multi-vendor federation)
    \item Impact: \textbf{High} (data loss, compliance breach, reputational damage)
\end{itemize}

\subsection{Case Study 2: Distributed Data Analysis}\label{case-study-2}
Here, A2A Clients in different departments analyze sensitive datasets by launching A2A Tasks to a central A2A Server, aggregating results via Artifacts.
\begin{enumerate}
    \item Layer 1: Foundation Models — Model inversion attacks may occur when adversarial input is embedded in A2A Message Parts, allowing attackers to extract sensitive data from LLMs.
    \item Layer 2: Data Operations — Data poisoning in A2A Artifacts, unauthorized aggregation of Task results, tool poisoning,  agent card poisoning, or tampering with distributed Task state are key threats.
    \item Layer 3: Agent Frameworks — Susceptible to Task replay via tasks/send, Agent Card spoofing, and schema violations in Task or Message objects.
    \item Layer 4: Deployment \& Infrastructure — Risks include network eavesdropping on A2A Task traffic and A2A Server compromise.
    \item Layer 5: Evaluation \& Observability — Insufficient anomaly detection in Task audit logs or log forgery in Task status events can enable undetected attacks.
    \item Layer 6: Security \& Compliance — Data privacy violations due to misconfigured Agent Cards or weak encryption of Task data.
    \item Layer 7: Agent Ecosystem — Data silo bridging via Agent Card enumeration and policy conflicts between federated A2A Servers can lead to unauthorized data flows.
\end{enumerate}
\subsubsection{Cross-Layer Vulnerabilities:}
\begin{itemize}
    \item Poisoned A2A Artifacts corrupting analytics (Layer 2 $\rightarrow$ Layer 1)
    \item Weak Task orchestration enabling replay or hijacking (Layer 3 $\rightarrow$ Layer 4)
    \item Cross-department Agent Card trust failures (Layer 7 $\rightarrow$ Layer 6)
\end{itemize}

\subsubsection{Risk Assessment}
\begin{itemize}
    \item Likelihood: \textbf{Medium-High} (complexity, distributed trust)
    \item Impact: \textbf{High} (business intelligence compromise, regulatory risk)
\end{itemize}

\subsection{Threat Evolution}
Threats to A2A-based multi-agent systems evolve as the protocol, Agent Card registry, and deployment patterns change. Regular threat modeling via the MAESTRO methodology, and updates to identified threats, are essential to address new attack techniques and changes in the A2A ecosystem.

\section{Mitigating Security Threats in A2A-Based Multi-Agent Systems} \label{mitigating-security-threats}

Drawing from the MAESTRO threat modeling methodology, Section-\ref{sec:threats} outlined several potential threats facing A2A-based MAS. This section delves into specific security controls and best practices to address these threats, as well as additional security considerations from Section-\ref{threats:addition sec}. Moreover, it re-contextualizes the case studies of Sections-\ref{case-study-1} and \ref{case-study-2} in light of those mitigations.

\subsection{Addressing Specific Threats and Enhancing Security}

The following subsections detail mitigation strategies for each threat identified in Section-\ref{sec:threats}, incorporating the additional considerations of Section-\ref{threats:addition sec}:

\subsubsection{Addressing Agent Card Spoofing} Addressing Section-\ref{agent-card-spoofing}
Agent Card Spoofing, where an attacker publishes a forged /.well-known/agent.json (Agent Card) at a malicious domain, poses a significant risk of task hijacking, data exfiltration, and agent impersonation.

\textbf{Mitigation Strategies:}
\begin{itemize}
    \item \textbf{Digital Signatures on Agent Cards:} Use digital signatures via a trusted Certificate Authority (CA) to ensure authenticity and integrity.
    \item \textbf{Secure Agent Card Resolution:} Use HTTPS with certificate validation and optionally, certificate pinning.
    \item \textbf{Agent Card Registry and Validation:} Use a trusted registry or directory for validation.
    \item \textbf{Reputation-Based Trust:} Implement a reputation system for rating Agent Cards.
    \item \textbf{Agent Card Sanitization:} Sanitize AgentCard content before using it with Foundational Models.
\end{itemize}

\subsubsection{Preventing A2A Task Replay} Addressing Section-\ref{a2a-task-replay}

\textbf{Mitigation Strategies:}
\begin{itemize}
    \item Include a unique \textbf{nonce} in each tasks/send request.
    \item Use \textbf{timestamp verification} with an acceptable time window.
    \item Use \textbf{Message Authentication Codes (MACs)}.
    \item Design tasks to be \textbf{idempotent}.
    \item Implement \textbf{session management} to track tasks.
\end{itemize}

\subsubsection{Preventing A2A Message Schema Violations} Addressing Section-\ref{a2a-schema-violation}

\textbf{Mitigation Strategies:}
\begin{itemize}
    \item Enforce \textbf{strict schema validation}.
    \item Sanitize all input from clients.
    \item Use \textbf{Content Security Policies (CSP)}.
    \item Execute tasks with \textbf{least privilege}.
\end{itemize}

\subsubsection{Preventing A2A Server Impersonation } Addressing Section-\ref{a2a-server-impersonation}

\textbf{Mitigation Strategies:}
\begin{itemize}
    \item Use \textbf{Mutual TLS (mTLS)}.
    \item Use \textbf{DNSSEC} to protect against spoofing.
    \item Implement \textbf{certificate pinning}.
    \item Apply \textbf{monitoring and intrusion detection}.
\end{itemize}

\subsubsection{Preventing Cross-Agent Task Escalation} Addressing Section-\ref{cross-agent-task-esc}

\textbf{Mitigation Strategies:}
\begin{itemize}
    \item Enforce strict \textbf{authentication and authorization}.
    \item \textbf{Validate credentials} for every task.
    \item Implement \textbf{audit logging}.
    \item Follow \textbf{least privilege} principles.
    \item Use \textbf{secure discovery} for Agent Cards.
\end{itemize}

\subsubsection{Mitigating Artifact Tampering via A2A Artifacts} Addressing Section-\ref{a2a-artifact-tampering}

\textbf{Mitigation Strategies:}
\begin{itemize}
    \item Use \textbf{digital signatures} on artifacts.
    \item Apply \textbf{hashing and checksums}.
    \item Use \textbf{encryption}.
    \item Ensure \textbf{secure storage and transmission}.
\end{itemize}

\subsubsection{Preventing Insider Threat/Logging Evasion via A2A Task Manipulation} Addressing Section-\ref{insider-threat}

\textbf{Mitigation Strategies:}
\begin{itemize}
    \item Implement \textbf{RBAC}.
    \item Apply \textbf{audit logging with integrity checks}.
    \item Ensure \textbf{separation of duties}.
    \item Conduct \textbf{security audits}.
    \item Require \textbf{MFA} for privileged roles.
\end{itemize}

\subsubsection{Addressing Supply Chain Attacks via A2A Dependencies} Addressing Section-\ref{supply-chain}

\textbf{Mitigation Strategies:}
\begin{itemize}
    \item Use \textbf{dependency scanning}.
    \item Apply \textbf{dependency pinning}.
    \item Generate and maintain a \textbf{Software Bill of Materials (SBOM)}.
    \item Conduct \textbf{vendor security assessments}.
    \item Follow \textbf{secure development practices}.
\end{itemize}

\subsubsection{Mitigating Authentication \& Identity Threats} Addressing Section-\ref{auth-identity-threats}

\textbf{Mitigation Strategies:}
\begin{itemize}
    \item Apply \textbf{strong JWT validation}.
    \item Use \textbf{secure token storage}.
    \item Implement \textbf{token rotation}.
    \item Use \textbf{mTLS for API access}.
    \item Follow \textbf{OAuth 2.0 best practices}, including PKCE.
\end{itemize}

\subsubsection{Mitigating Poisoned AgentCard} Addressing Section \ref{poisioned-agetnt-card}

\textbf{Mitigation Strategies:}
\begin{itemize}
    \item Apply \textbf{input sanitization} to all AgentCard content.
    \item Use a \textbf{whitelist of allowed characters}.
    \item Implement \textbf{escaping/encoding} of special characters.
    \item Enforce \textbf{Content Security Policy (CSP)}.
    \item Validate structure with \textbf{schema checks and type constraints}.
\end{itemize}

\subsection{Additional Security Considerations} Addressing Section-\ref{threats:addition sec}
\begin{itemize}
    \item Continuously scan and verify supply chain dependencies.
    \item Develop and maintain incident response and recovery plans.
    \item Evaluate security posture of open-source libraries.
    \item Use declarative security configurations (YAML/JSON).
    \item Integrate security testing into CI/CD pipelines.
    \item Document capabilities and trust boundaries in Agent Cards.
\end{itemize}

\subsection{Applying Mitigations to Case Studies (Sections \ref{case-study-1} and \ref{case-study-2})}

\subsubsection{Case Study 1: Collaborative Document Processing}
\begin{itemize}
    \item Digitally sign all documents.
    \item Enforce granular access control.
    \item Apply DLP techniques.
    \item Sanitize Agent Cards before using with FMs.
    \item Validate and authenticate all task submissions.
\end{itemize}

\subsubsection{Case Study 2: Distributed Data Analysis}
\begin{itemize}
    \item Implement differential privacy.
    \item Use federated learning.
    \item Apply secure multi-party computation (SMPC).
    \item Sanitize Agent Cards before use.
    \item Apply strict audit controls on artifact aggregation.
\end{itemize}

\section{Secure Application Development Strategies}

\subsection{Securely Implementing with Current A2A Features}
This involves endpoint hardening, augmenting authentication mechanisms beyond the basic requirements if necessary, and rigorous input/output validation.

\subsection{Leveraging Enhanced A2A Features for Advanced Security}
Future enhancements or complementary technologies could include DID-based authentication, granular policy enforcement frameworks, and mechanisms for establishing data provenance.

\section{Key Control Measures for Safe A2A Deployment}
This section covers essential controls for deploying A2A-based applications securely:
\begin{itemize}
    \item Endpoint security (TLS, network controls).
    \item Strong authentication and authorization.
    \item Comprehensive input validation and sanitization.
    \item Principle of least privilege for agent capabilities and data access.
    \item Robust monitoring, logging, and alerting.
    \item Secure software development lifecycle (SSDLC) practices.
    \item Incident response planning and execution.
\end{itemize}
    For secure A2A coding examples, please refer to the Github repository \footnote{\url{https://github.com/kenhuangus/a2a-secure-coding-examples} \label{fn:gh_repo}}.

\section{Implementing A2A Server Securely}

\begin{figure*}
    \centering
    \includegraphics[width=1\linewidth]{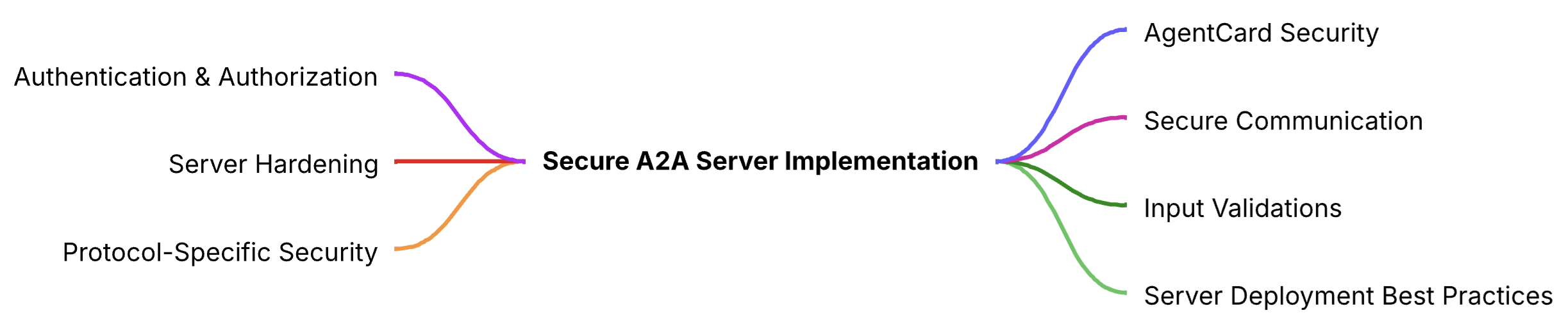}
    \caption{Best Practices For Secured A2A Server}
    \label{fig:a2aserver}
\end{figure*}

In this section, we will focus on suggesting specific security controls to implement for the A2A Server deployment, aiming to enhance its resilience against potential threats. A high-level overview of these threats is illustrated in Figure~\ref{fig:a2aserver}. 
\subsection{Securing the AgentCard}
The AgentCard (\texttt{/.well-known/agent.json}) is a critical security element as it exposes information about your agent's capabilities and authentication requirements.

\subsection{Securing the AgentCard}


The AgentCard is a critical security element in the A2A protocol as it exposes information about an agent's capabilities and authentication requirements.
\begin{enumerate}
    \item \textbf{AgentCard Location and Access Controls}
        \begin{itemize}
            \item Host the AgentCard file \texttt{(/.well-known/agent.json)} with appropriate access controls.
            \item Implement rate limiting to prevent enumeration attacks.
            \item Consider using content security headers to prevent unauthorized embedding or framing.
        \end{itemize}

    \item \textbf{AgentCard Content Security}
        \begin{itemize}
            \item Include only necessary information about your agent's capabilities.
            \item Specify detailed authentication requirements using OpenAPI 3.x Security Scheme objects.
            \item Validate the AgentCard content regularly to ensure it doesn't expose sensitive information.
            \item Keep authentication details accurate and up to date.
        \end{itemize}
\end{enumerate}
\subsection{Authentication and Authorization}
\begin{itemize}
    \item \textbf{Establish Server Identity:} Authenticate the server's identity via digital certificates provided by trustworthy Certificate Authorities. Employ TLS to secure connections, enabling clients to authenticate the server's authenticity during the handshake and mitigate man-in-the-middle attacks.
    \item \textbf{Declare Authentication Protocols:} Explicitly declare the supported authentication methods (e.g., OAuth, OIDC, API Keys) in the Agent Card.
    \item \textbf{Authenticate Each Request:} Require authentication for each incoming HTTP request. Ensure that every request has proper credentials (e.g., tokens, certificates). Reject requests that lack the valid credentials by utilizing suitable HTTP status codes (e.g., 401 Unauthorized, 403 Forbidden).
\end{itemize}

\begin{enumerate}
    \item \textbf{Protecting Sensitive Actions and Data} \\
    It is recommended that A2A server will protect sensitive information by managing authorization to both ‘skills’ and ‘tools’:
    \begin{itemize}
        \item \textbf{Skills:} Agents are required to advertise their skills through an Agent Card, showcasing their expertise. It is recommended for agents to grant permission for each skill or grant permission on a per-skill basis using specific scopes, enabling different access levels (e.g., read-only skills).
        \item \textbf{Tools:} Agents are required to limit access to sensitive data and actions by securing using controlled tools. Therefore, when agentic flow requests data, the agent will grant permissions based on this.
    \end{itemize}

    \item \textbf{API Keys} \\
    For simpler implementations, API keys may be used:
    \begin{itemize}
        \item Generate strong, random API keys following cryptographic best practices.
        \item Implement key rotation policies to regularly update API keys.
        \item Store API keys securely and never expose them in client-side code.
    \end{itemize}

    \item \textbf{JWT Authentication} \\
    For stateless authentication:
    \begin{itemize}
        \item Implement robust JWT validation including signature verification.
        \item Set appropriate token expiration times to limit the impact of token theft.
        \item Include only necessary claims in the JWT payload.
    \end{itemize}
\end{enumerate}

\subsection{Secure Communication}
\begin{enumerate}
    \item \textbf{Transport Layer Security (TLS)}
    \begin{itemize}
        \item Enforce HTTPS for all A2A communications with proper TLS configuration (TLS1.3).
        \item Regularly renew TLS certificates and disable insecure ciphers.
    \end{itemize}

    \item \textbf{Data Protection in Transit}
    \begin{itemize}
        \item Ensure all A2A messages are encrypted during transmission.
        \item Validate certificate chains to prevent man-in-the-middle attacks.
        \item Consider implementing certificate pinning for critical connections.
    \end{itemize}

    \item \textbf{Data Protection at Rest}
    \begin{itemize}
        \item Encrypt sensitive data stored by your A2A server.
        \item Secure storage of any persistent data from agent interactions.
        \item Implement proper key management for encryption keys.
    \end{itemize}
\end{enumerate}

\subsection{Input Validation and Request Processing}
\begin{enumerate}
    \item \textbf{Message Validation}
    \begin{itemize}
        \item Validate all incoming messages against the A2A protocol schema.
        \item Implement robust input sanitization to prevent injection attacks.
        \item Verify message formats, sizes, and content types before processing.
    \end{itemize}

    \item \textbf{URI Validation}
    \begin{itemize}
        \item Strictly validate any URIs included in messages to prevent Server-Side Request Forgery (SSRF) attacks.
        \item Implement allow-lists for acceptable URI schemes and domains.
        \item Avoid fetching content from untrusted or user-supplied URIs.
    \end{itemize}

    \item \textbf{Processing File Parts} \\
    When handling \texttt{FilePart} content in A2A messages:
    \begin{itemize}
        \item Scan all uploaded files for malware.
        \item Validate file types and enforce size limits.
        \item Store uploaded files securely with appropriate access controls.
    \end{itemize}
\end{enumerate}

\subsection{Server Implementation Best Practices}
\begin{enumerate}
    \item \textbf{Error Handling and Logging}
    \begin{itemize}
        \item Implement comprehensive error handling that doesn't expose sensitive information.
        \item Maintain detailed security logs for authentication attempts, authorization decisions, and security events.
        \item Ensure logs are protected and cannot be tampered with.
        \item Implement monitoring and alerting for suspicious activities.
    \end{itemize}

    \item \textbf{Rate Limiting and DoS Protection}
    \begin{itemize}
        \item Implement rate limiting on all A2A endpoints to prevent abuse.
        \item Consider using exponential backoff for failed authentication attempts.
        \item Monitor for and mitigate denial of service attacks targeting your A2A server.
    \end{itemize}

    \item \textbf{Secure Development Practices}
    \begin{itemize}
        \item Follow secure coding guidelines specific to your implementation language.
        \item Conduct regular security code reviews.
        \item Implement application security testing as part of your development workflow.
        \item Keep all dependencies updated to address security vulnerabilities.
    \end{itemize}
\end{enumerate}
\begin{table*}[htbp]
  \centering
  \caption{Comparative Analysis of A2A and MCP}
  \label{tab:a2a_mcp_comparison}
  \begin{tabular}{@{}lll@{}}
    \toprule
    \textbf{Feature} & \textbf{Google Agent2Agent (A2A)} & \textbf{Anthropic Model Context Protocol (MCP)} \\
    \midrule
    \textbf{Purpose} & Enable interoperability between diverse AI agents & \makecell[l]{Standardize connection between AI models/agents \\ and external tools/data} \\
    \textbf{Focus} & \makecell[l]{Agent-to-Agent collaboration, delegation, \\ messaging} & \makecell[l]{Agent-to-Tool/Resource access, context \\ provisioning} \\
    \textbf{Primary Interaction} & Client Agent $\leftrightarrow$ Remote Agent & \makecell[l]{MCP Client (Agent/Host) $\leftrightarrow$ MCP Server \\ (Tool/Data)} \\
    \textbf{Key Mechanisms} & \makecell[l]{Agent Cards (discovery), Task object (lifecycle), \\ Messages, Artifacts} & \makecell[l]{Tools, Resources, Prompts (exposed by server), \\ Client-Server requests} \\
    \textbf{Ecosystem Role} & \makecell[l]{Horizontal Integration \\ (Agent Network Communication)} & \makecell[l]{Vertical Integration \\ (Agent Capability Enhancement)} \\
    \bottomrule
  \end{tabular}
\end{table*}

\subsection{Protocol-Specific Security Considerations}
\begin{enumerate}
    \item \textbf{Streaming and SSE (Server-Sent Events)} \\
    For implementations using tasks/sendSubscribe and Server-Sent Events (SSE):
    \begin{itemize}
        \item Implement proper authentication for SSE connections.
        \item Maintain secure state management for long-lived connections.
        \item Monitor for and mitigate resource exhaustion attacks.
    \end{itemize}

    \item \textbf{Push Notifications} \\
    When implementing the optional push notifications capability:
    \begin{itemize}
        \item Validate webhook URLs rigorously to prevent SSRF attacks.
        \item Implement signature verification for webhook payloads.
        \item Use HTTPS for all webhook communications.
        \item Implement retry policies with appropriate backoff.
    \end{itemize}

    \item \textbf{Connection Management and Abuse Prevention} \\
    Effective connection management is essential to ensure the scalability and resilience of A2A Server streaming features such as \texttt{tasks/sendSubscribe} and Server-Sent Events (SSE). To prevent abuse and resource exhaustion, the server should enforce various strategies depicted below.

    \item \textbf{Connection Quotas:} connection quotas per client or IP, close idle connections through timeout mechanisms, and use periodic keep-alive pings to detect and clean up stale sessions.

    \item \textbf{Connection limits:} Set hard limits on the number of active SSE connections per client ID/IP to prevent resource exhaustion.

    \item \textbf{Idle Timeout \& Keep-Alive Pings:} Enforce connection idle timeouts. Use periodic keep-alive pings and terminate stale sessions.

    \item \textbf{Backpressure-Aware Streaming:} Drop non-critical event messages for lagging clients or apply backoff strategies to prevent memory build-up.
\end{enumerate}

\subsection{Server Hardening}
When deploying A2A server, it is essential to use a hardened environment following security best practices.
\begin{itemize}
    \item Implement network segmentation to isolate the A2A server.
    \item Use Web Application Firewalls (WAF) to protect against common web attacks.
    \item Regularly apply security updates to server components.
    \item Implement comprehensive monitoring of your A2A server.
    \item Develop incident response procedures for security breaches.
    \item Conduct regular security assessments and penetration tests.
    \item Establish a security incident management process.
\end{itemize}

\begin{figure*}[tbp] 
    \centering
    \includegraphics[width=0.9\linewidth]{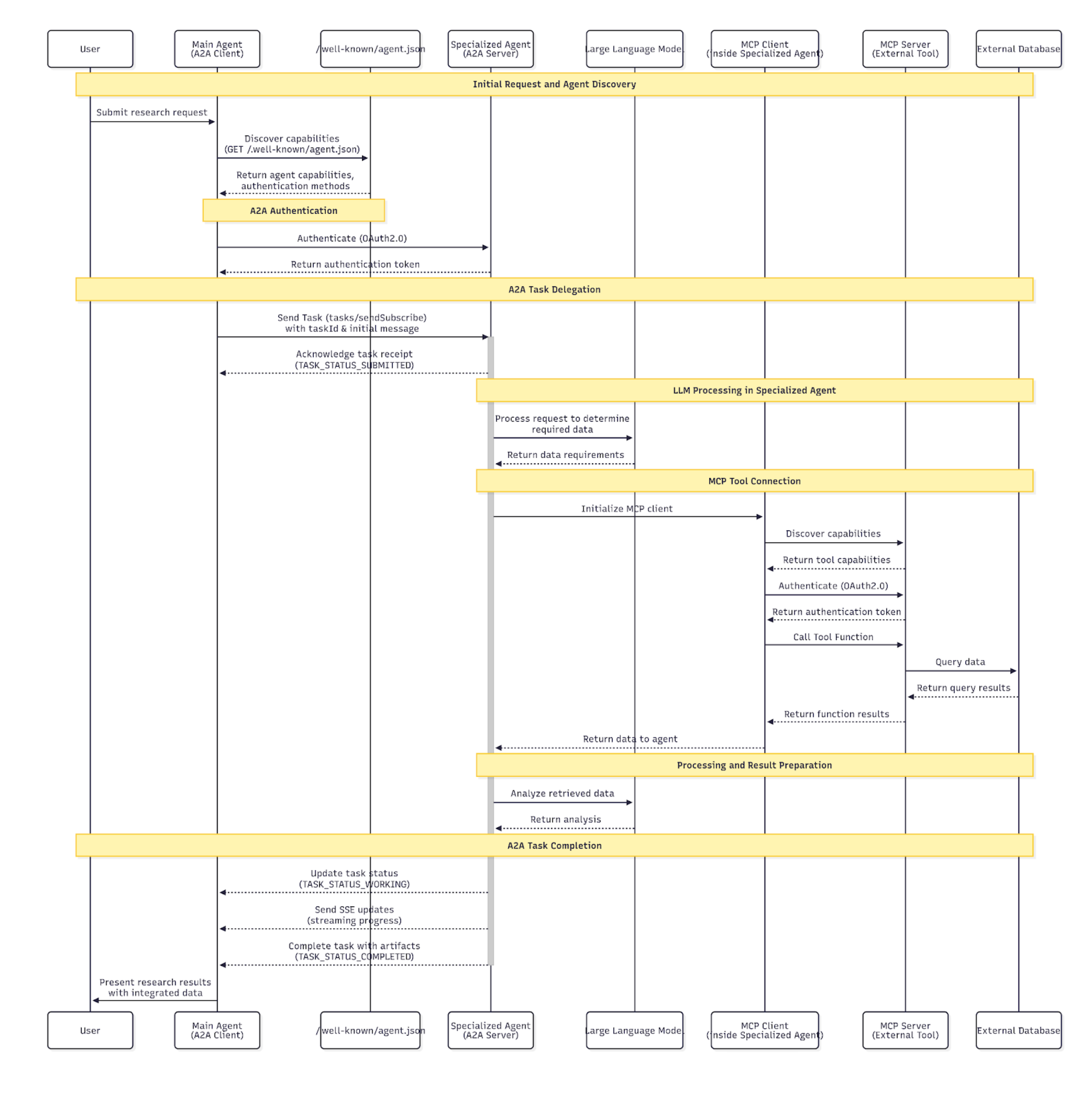} 
    \caption{End to End Agents collaboration utilizing A2A and MCP}
    \label{fig:flow}
\end{figure*}
\section{MCP and A2A: A Synergy}
The Agent-to-Agent (A2A) protocol and Model Context Protocol (MCP) represent complementary frameworks that together create a robust foundation for sophisticated agentic systems. Rather than competing standards, these protocols operate at different layers of the AI interaction stack, enabling both horizontal coordination between peer agents and vertical integration with specialized tools and data sources. Table~\ref{tab:a2a_mcp_comparison} presents a comparative analysis of these two critical protocols in Agentic AI.

 When deployed together, these protocols create an efficient hierarchical workflow system. An agent initially receives a complex task from either a human user or another agent through A2A protocols. To successfully complete this assignment, the agent often needs specialized capabilities beyond its own scope. Using A2A's discovery mechanism, it identifies and delegates specific subtasks to purpose-built agents with relevant expertise, such as Claim Agents or Rental Car Agents. These specialized agents then leverage MCP to connect with structured data sources, computational tools, or external systems required to fulfill their assigned responsibilities. The completed work flows back through the agent hierarchy via A2A's structured task management framework, enabling seamless integration of results from multiple specialized systems into a cohesive solution delivered to the original requestor. Figure~\ref{fig:flow} illustrates the steps in such a use case. 
 
 The flow between a Claim Agent and Rental Car Agent would utilize the A2A protocol for agent-to-agent coordination and task management, where the Claim Agent (acting as the Main A2A Client) would discover the Rental Car Agent's capabilities through the well-known agent registry, authenticate, and delegate specific rental-related tasks with unique taskIds. Meanwhile, the MCP framework enables each specialized agent to extend its capabilities by connecting to external tools and databases - the Claim Agent might access policy databases while the Rental Car Agent connects to vehicle availability systems. These frameworks complement each other by allowing the Claim Agent to maintain high-level coordination of the insurance claim process (tracking status updates and compiling final results for the user) while the Rental Car Agent uses MCP to perform specialized functions like querying vehicle inventories and processing rental agreements, ultimately delivering an integrated solution where claim processing and rental car arrangements happen seamlessly within a unified workflow.

This architectural approach promotes several key design principles. It enables functional modularity, allowing organizations to develop specialized agents with deep domain expertise. It supports compositional flexibility, as different agent combinations can be assembled to address diverse problems. It maintains clear separation of concerns between peer-level agent coordination and tool-level resource access. Most importantly, it creates an extensible ecosystem where new capabilities can be added incrementally without redesigning the entire system.

The multi-protocol architecture introduces important security considerations at integration boundaries. Potential attack vectors may cross protocol boundaries—for example, an authentication vulnerability in the A2A layer could be exploited to gain unauthorized MCP access, while sensitive information exposed through inadequately protected MCP connections might enable subsequent A2A-based compromises. Therefore, robust security strategies must address not only each protocol individually but also the critical interfaces where they interconnect. This comprehensive approach ensures continuous protection throughout the entire agent workflow chain, with special attention to authentication token handling, task delegation permissions, and secure data transmission between the Claim Agent, Rental Car Agent, and their respective external systems. \cite{narajala2025enterprise}

\section{Conclusion}

\subsection{Recap of A2A's Role, Security Challenges, and Enhancement Potential}
The Agent-to-Agent (A2A) protocol plays a foundational role in enabling secure, composable, and scalable multi-agent systems. By standardizing how agents discover, authenticate, and communicate with one another, A2A facilitates a new paradigm of autonomous collaboration—ranging from document co-authoring agents to federated analytics across distributed datasets. Its declarative nature and emphasis on explicit capabilities (via AgentCards) empower developers to build trust-aware, modular workflows that reflect real-world boundaries and roles.
However, with these capabilities come significant security challenges. The distributed and dynamic nature of multi-agent systems introduces threats. Furthermore, novel risks such as prompt injection into AgentCards and cross-agent escalation highlight the need for context-aware threat modeling—such as that provided by the MAESTRO framework.
To realize the full potential of A2A-based systems, security must be deeply integrated into both protocol design and server implementation. From strong cryptographic controls and zero-trust agent authentication to schema validation, logging integrity, and resilient streaming protocols, robust security controls are essential. As multi-agent ecosystems grow in complexity, future enhancements to A2A should emphasize adaptive trust, continuous policy enforcement, and secure-by-default configurations. By proactively addressing today’s threats and anticipating tomorrow’s, A2A can serve as a resilient backbone for intelligent, secure, and trustworthy agentic AI applications.

\subsection{Future Directions for Secure Agent Collaboration}
Looking ahead, the secure collaboration of autonomous agents hinges on continued progress in protocol standardization and the widespread adoption of security best practices. Initiatives like A2A and MCP are crucial, but their evolution must prioritize security considerations, potentially incorporating lessons from Zero Trust architectures adapted for agent-specific contexts. Future work should focus on developing standardized methods for expressing and enforcing complex authorization policies between agents, establishing robust mechanisms for agent identity verification and reputation management, and creating frameworks for secure multi-agent coordination, especially when crossing administrative or trust boundaries. Furthermore, addressing the security implications of integrating Large Language Models within agentic workflows and ensuring the resilience of agent communication against sophisticated attacks will be paramount for building trustworthy and reliable multi-agent systems. The development of comprehensive threat models, like MAESTRO \cite{Huang2025Maestro}, and shared security implementation guidelines will be vital for guiding developers in building the next generation of secure agentic AI applications.

\bibliographystyle{IEEEtran} 
\bibliography{references} 

\end{document}